\documentclass[aps,prb,groupedaddress,floats,reprint]{revtex4-2}

\usepackage[german,english]{babel}

\usepackage{amsfonts,graphicx,soul,natbib,mathrsfs}
\usepackage[dvipsnames]{xcolor}

\usepackage[normalem]{ulem}  

\usepackage[colorlinks=true]{hyperref}

\usepackage{epstopdf}
\usepackage{textcomp}
\usepackage{xcolor}
\usepackage{soul}
\usepackage{physics}
\usepackage{graphicx}
\usepackage{lipsum}
\usepackage{comment}
\usepackage{amssymb}

\newcommand{\expv}[1]{\left\langle #1 \right\rangle}

\usepackage{bm}

\usepackage{amsmath,amssymb,amsfonts}
\usepackage{fancyhdr}
\usepackage{lipsum}
\usepackage{subcaption}
\usepackage{multirow}
\usepackage{dcolumn}
\usepackage{bm}

\begin{document}

\title{Elastic scattering of Airy electron packets on atoms}
\author{D. Grosman}%
\affiliation{School of Physics and Engineering,
ITMO University, 197101 St. Petersburg, Russia}%

\author{N. Sheremet}%
\affiliation{School of Physics and Engineering,
ITMO University, 197101 St. Petersburg, Russia}%

\author{I. Pavlov}%
\affiliation{School of Physics and Engineering,
ITMO University, 197101 St. Petersburg, Russia}%

\author{D. Karlovets}%
\affiliation{School of Physics and Engineering,
ITMO University, 197101 St. Petersburg, Russia}

\date{\today}
\begin{abstract}
    The problem of elastic scattering of the non-relativistic electron Airy beams on potential fields is considered for a hydrogen atom in the ground 1s state. 
    It is demonstrated that  the angular dependence of the scattering probability density is in general azimuthally asymmetric. When the position of the atom happens to coincide with one the minima of the probability density of the Airy beam, the asymmetric pattern is represented by four separated peaks. We show that this behaviour is very sensitive to the precision with which the relative position of the atom and the minima is defined and study how uncertainty in the position measured in terms of the wave-packet width affects observation of the azimuthal asymmetry. Finally, we consider a spatially localized target and discuss the difficulties of observing the asymmetry for targets with sizes exceeding the critical value determined by the beam parameters and by the position of the target center.
\end{abstract}

\maketitle

\section{Introduction}
In this paper we aim to give a detailed analysis of elastic scattering of an electron wave packet of a special shape by a potential field. Real particles are localized both in the coordinate and momentum space, which results in nonuniform densities and imposes the necessity of describing them as wave-packets, however in a majority of scenarios they are represented as plane waves, which is valid when the characteristic distances in the problem considered are much smaller than the typical sizes of inhomogeneity, i.e, the distances at which the nonuniform character of the density comes into play. Nevertheless, there are examples of experimentally investigated scattering processes where macroscopically large impact parameters greatly contributed to the cross section \cite{KOTKIN1992}. In this case plane-wave description is no longer sufficient and particles must be represented as wave-packets.

Non-Gaussian wave-packet solutions to the Shr\"odinger equation that are of special interest due to the presence of phase have been known for over several decades now. However, detailed analysis for massive particles (say, electrons) is still rather scarce. In 2010-2011 it was reported on experimental generation of vortex electron states carrying definite value \textit{$\hbar l$}   of \textit{intrinsic angular orbital momentum} along the particle propagation axis where the orbital quantum number \textit{$l$} can already be as high as one thousand \cite{Uchida2010,McMorran2011}. These states are characterized by spiraling current density proportional to the orbital quantum number $l$ due to azimuthal dependence of the wave function. Later, the electron Airy beams with energies up to 200 keV were experimentally created by diffraction of Gaussian electrons on nano-scale holograms \cite{VolochBloch2013}. The distinguishing feature of Airy beams is the cubic dependence of the phase of the wave function on the particle momentum $\varphi(\bm{k}) \sim \xi_x^3 k_x^3 + \xi_y^3 k_y^3$ with $\bm{\xi}=\{\xi_x, \xi_y\}$ being a 2D vector which parameterizes Airy beams and transforms as coordinates under the Lorentz boosts \cite{Karlovets2017a}. \textcolor{black}{In the limit $\xi_{x, y} \to 0$ Airy packet coincides with a Gaussian one, while for $\xi_{x, y}$ much larger than the size of the Gaussian packet it tends to a non square integrable packet \cite{Berry}. Among the features of the Airy packet can be identified the presence of intrinsic quadrupole moment \cite{Karlovets2019}.} Currently, electron Airy beams are not used as widely as their optical counterparts. However, photon  Airy beams have found various applications in optical micromanipulation \cite{Baumgartl2008}, optical trapping \cite{Zheng2011,Zhang2011}, generation of plasma channels \cite{Polynkin2009}, surface Airy plasmons \cite{Salandrino2010,Minovich2011} and applications in lasers \cite{Porat2011,Longhi2011}.


\textcolor{black}{Up to the present, most of the structured electron collision studies treated twisted electrons. Scattering of free twisted particles has been researched in \cite{Ivanov2011} and \cite{Ivanov_Coulomb}. Inelastic \cite{ VanBoxem2015} and elastic \cite{VanBoxem2014, Karlovets2017} scattering of vortex non-relativistic electrons by atomic targets has also been considered in detail. Interaction of twisted relativistic electrons with atomic targets has been examined both in the Born approximation \cite{Serbo2015} and beyond it \cite{Kosheleva2018}. More complex processes such as ionization of atomic targets using electron vortex beam projectiles  \cite{Harris_2019, Plumadore_2020, Dhankhar_2020} and angular momentum transfer \cite{LLoyd2012_Dichroism, Lloyd2012} have been thoroughly investigated. In the paper \cite{Karlovets2015} the well-known Born approximation was generalized for the case when the incident beam is a wave packet of a \textit{general form}. In the present paper we aim to apply this theory to the scattering of Airy electron packets on an atomic potential.}




The structure of the paper is as follows. In Sec.II A. we recall the standard Born approximation including the well known formulas. Then we discuss generalization of these formulas for scattering of wave-packets developed in \cite{Karlovets2015} and present the expression for the number of events. In Sec. II B. we consider special case of the Airy beam and derive the scattering amplitude by making use of the theory presented in previous sections. Then in Sec. II C. and D. we consider the generalization of the number of events for scattering on a macroscopic (infinitely wide) and mesoscopic (localized) target respectively. In Sec. III we present density plots for different scattering scenarios and introduce special points of the first and second type as well as transitional points which are characterized by distinctive scattering patterns. We analyze the feasibility of the scattering patterns and their sensitivity to the parameters of the beam and scatterer. Then we discuss the azimuthal dependence of the number of events and finally study how the scattering patterns alter when switching to consideration of mesoscopic targets.

\section{Theoretical background}
\subsection{Basic formulas}
Consider the problem of scattering of a charged non-relativistic particle (electron) off a spherically symmetric potential field $U(r)$ with a typical radius of action $a$, which will be the Bohr radius in what follows. For the initial and scattered electrons being plane waves, the S-matrix element for the transition between states with momentum $\bm{p}_i$ and $\bm{p}_f$ is expressed via the scattering amplitude $f(\varepsilon_i, \theta, \varphi)$
\begin{equation}
\begin{aligned}
    S_{fi}^{(pw)} = \bra{f}&S\ket{i} = (2\pi)^2 i \delta(\varepsilon_i - \varepsilon_f)\frac{f(\varepsilon_i,\theta,\varphi)}{m_e},\\
    &\varepsilon_i = \frac{\bm{p}_i^2}{2m_e}, \; \varepsilon_f = \frac{\bm{p}_f^2}{2m_e},
\end{aligned}
\end{equation}
where $m_e$ is the electron mass, and the corresponding number of scattering events and the differential cross section are
\begin{equation}
\begin{aligned}
    &d\nu = N_e |\bra{f}S\ket{i}|^2\frac{d^3p_f}{(2\pi)^3}, \;\frac{d\sigma}{d\Omega} = |f(\varepsilon_i, \theta, \varphi)|^2.
\end{aligned}
\end{equation}
Here $N_e$ is the number of incident electrons.

For the initial state being a wave packet
\begin{equation}
    \ket{i} = \int \ket{\bm{k}} \Phi(\bm{k})\frac{d^3k}{(2\pi)^{3/2}},
\end{equation}
the matrix element for the transition into the plane wave state with momentum $\bm{p}_f$ is given by integration of the plane wave amplitudes with the packet's wave function in momentum representation
\begin{equation}
    S_{fi} = \int \bra{\bm{p}_f} S \ket{\bm{k}} \Phi(\bm{k}) \frac{d^3k}{(2\pi)^{3/2}} = \int S_{fk}^{(pw)} \Phi(\bm{k}) \frac{d^3k}{(2\pi)^{3/2}}.
\end{equation}
We will consider a wave packet propagating along the $z$ - direction on average
\begin{equation}
    \expv{\bm{k}} = (0,0,p_i)
\end{equation}
with non-zero average of the absolute value of the transverse momentum
\begin{equation}
    \expv{k_{\perp}} = \expv{|\bm{k}_{\perp}|} = \kappa_0 = p_i \tan\theta_k.
\end{equation}
Here $\theta_k$ is the conical angle. We assume no coupling of the longitudinal and transverse directions and hence factorization of the wave function into transverse and longitudinal parts
\begin{equation}
    \Phi(\bm{k}) = \Phi_{\perp}(\bm{k}_{\perp})\Phi_{long}(k_z).
\end{equation}
We also assume the dispersions to be small compared to the longitudinal momentum \cite{Hwang2014, Sun2004}:
\begin{equation}
        \Delta k_x = \Delta k_y \sim \frac{1}{\sigma_{\perp}} \ll p_i, \; \Delta k_z \sim \frac{1}{\sigma_z} \ll p_i,
\end{equation}
where $\sigma_{\perp}$ and $\sigma_{z}$ are the transverse and longitudinal averaged sizes of the electron packet. From the experimental point of view the interesting case is when the packet's size $\sigma_z$ is greater than the field's radius of action and yet still small enough to neglect the packet's spreading in the transverse plane during the collision:
\begin{equation}
\label{eq:size}
    a \ll \sigma_z \ll \sigma_{\perp}\frac{p_i}{\kappa_0}.
\end{equation}
Provided \eqref{eq:size} is fulfilled one can derive the following expression for the scattering amplitude \cite{Karlovets2015}:
\begin{equation}
\label{eq:NoE}
\begin{aligned}
    &\frac{d\nu}{d\Omega} = \frac{N_e}{\cos\theta_k}|F(\bm{Q})|^2,\\
    &F(\bm{Q}) = \int f(\bm{Q} - \bm{k}_{\perp}) \Phi_{\perp}(\bm{k}_{\perp})\frac{d^2k_{\perp}}{(2\pi)^2},\\
    &\bm{Q} = (p_f\sin\theta\cos\varphi, p_f\sin\theta\sin\varphi,p_f\cos\theta - p_i),
\end{aligned}
\end{equation}
where $p_f = \sqrt{p_i^2 + \kappa_0^2}$.

\subsection{Scattering of the Airy packet}
\begin{figure}[b]
\center{\includegraphics[width=1\linewidth]{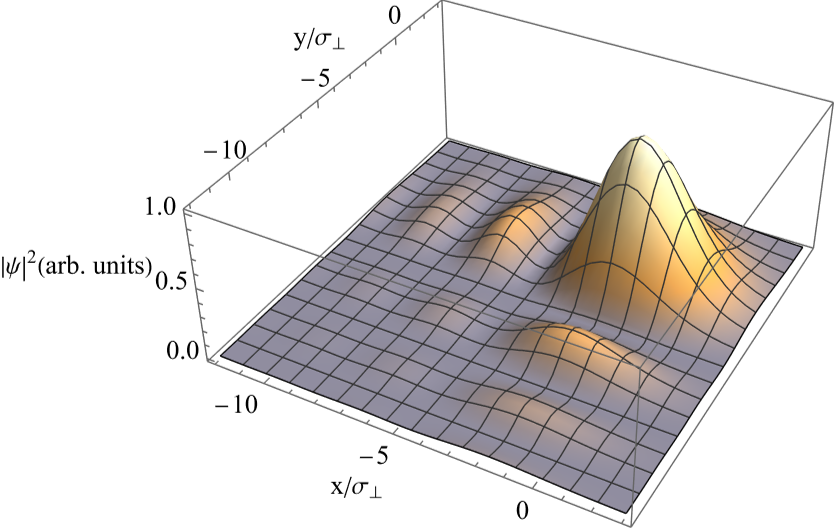}}
\caption{Spatial distribution of the Airy wave packet. $\xi_x=\xi_y=2\sigma_{\perp}$, $b_x=b_y=0$.}
\label{Fig:1}
\end{figure}
In this section we consider the special case of a wave packet with a phase in the momentum space ---  so-called Airy packet \cite{VolochBloch2013, Karlovets2017a}, whose transverse wave function in the
momentum representation is
\begin{equation}
\label{eq:airy}
\begin{aligned}
    \Phi_{\perp}(\bm{k}_{\perp}) &= N \exp\left(-k_x^2\sigma_{\perp}^2-k_y^2\sigma_{\perp}^2\right)\\
    &\exp(\frac{i}{3}\xi_x^3k_x^3+\frac{i}{3}\xi_y^3k_y^3 -i k_x b_x -i k_y b_y),
\end{aligned}
\end{equation}
where $N$ is the normalization constant defined by $\int |\Phi_{\perp}(\bm{k}_{\perp})|^2\frac{d^2k}{2\pi} = 1$ and $\bm{b} = \{b_x,b_y\}$ is the impact parameter introduced to account for non head-on collision scenarios. The probability density in coordinate representation is shown in Fig.\ref{Fig:1}.

We will consider two potentials - Hydrogen-like atom and Yukawa potential
\begin{equation}
    \begin{aligned}
        &U_{\text{hyd}}(r) = -\frac{e^2}{r}\left(1 + \frac{r}{a}\right)\exp(-\frac{2r}{a}),\\
        &U_{\text{Yu}}(r) = \frac{V_0}{r}\exp(-\mu r).
    \end{aligned}
\end{equation}
Here 
$\mu$ is the inverse effective radius of action of a Yukawa potential, $V_0$ is the effective Yukawa potential amplitude.
The corresponding Born amplitudes for the potentials are
\begin{equation}
\begin{aligned}
    &f_{\text{hyd}}(\bm{q}) = \frac{a}{2}\left[\frac{1}{1+(q a/2)^2}+\frac{1}{\left(1+(q a/2)^2\right)^2}\right],\\
    &f_{\text{Yu}}(\bm{q}) = -\frac{2m_e V_0}{q^2 + \mu^2},
\end{aligned}
\end{equation}
which can be written in a general form with the help of the function $I(\eta, z)$ defined as follows
\begin{equation}
\begin{aligned}
    &I(\eta,z) = \int\limits_{0}^{\infty}(1 + \eta s)e^{-sz}ds = f_0 \left(\frac{1}{z} + \frac{\eta}{z^2}\right),\\
     &f_{\text{hyd}}(\bm{q}) = \frac{a}{2}I\left(1,1+\frac{q^2a^2}{4}\right),\\
    &f_{\text{Yu}}(\bm{q}) = -\frac{2m_eV_0}{\mu^2}I\left(0,1+\frac{4q^2}{\mu^2}\right).
\end{aligned}
\end{equation}
Here $f_0$ is the amplitude of the potentials which is either $f_0 =\frac{a}{2}$ or $f_0 = -\frac{2m_eV_0}{\mu^2}$.
The amplitude for scattering of an Airy packet on either one of the potentials of interest is then also written in a general form
\begin{equation}
\begin{aligned}
    &F(\bm{Q},\bm{b},\eta,a,f_0) = \\
    &f_0\int I\left(\eta, 1 + \frac{1}{4}(\bm{Q}-\bm{k}_{\perp})^2a^2\right)\Phi_{\perp}(\bm{k}_{\perp})\frac{d^2k_{\perp}}{(2\pi)^2}.
\end{aligned}
\end{equation}
Note that the function $F\left(\bm{Q},\bm{b},1,a,\frac{a}{2}\right)$ - is the amplitude for the scattering of an Airy packet on a ground state hydrogen atom into a plane wave state and $F\left(\bm{Q},\bm{b},0,\frac{2}{\mu},-\frac{2m_eV_0}{\mu^2}\right)$ - is the amplitude for the same process but involving Yukawa potential.
The final expression for the amplitude is given by a one dimensional integral
\begin{equation}
\label{eq:amp}
    \begin{aligned}
    &F(\bm{Q},\bm{b},\bm{\xi},\eta, a, f_0) = f_0 N\frac{(2\pi)^2}{\xi_{x}\xi_{y}} \\
    &\times\int\limits_{0}^{\infty}ds (1+\eta s)\exp\left\{-s\left(1+\frac{1}{4}Q^2a^2\right)\right\}\\
    &\times\exp\left\{\frac{2}{3}\rho_x^6(s)-i\rho_x^2(s)\zeta_{x}(s)+\frac{2}{3}\rho_y^6(s)-i\rho_y^2(s)\zeta_{y}(s)\right\}\\
    &\times \operatorname{Ai}\left[-i\zeta_{x}(s)+\rho_x^4(s)\right]\operatorname{Ai}\left[-i\zeta_{y}(s)+\rho_y^4(s)\right],
    \end{aligned}
\end{equation}
where the expressions 
\begin{equation}
\begin{aligned}
\label{eq:p}
    &\rho_{x,y}^2(s) = \frac{1}{\xi^2_{x,y}}\left(a^2s/4+\sigma_{\perp}^2\right),\\
    &\zeta_{x,y}(s) = \frac{Q_{x,y}a^2s}{2\xi_{x,y}} - i \frac{b_{x,y}}{\xi_{x,y}},
\end{aligned}
\end{equation}
will be given physical interpretation shortly.

Expression \eqref{eq:amp} describes both the hydrogen atom and Yukawa potential. In the formula for the scattering amplitude $\operatorname{Ai}(z)$ is the Airy function properly defined for complex arguments:
\begin{equation}
    \operatorname{Ai}(z) = \frac{1}{2\pi i}\int\limits_{C}\exp(zt -\frac{t^3}{3})dt, 
\end{equation}
where the integral is over a path $C$ starting at a point at infinity with $\frac{\pi}{2} < \text{arg(t)} < \frac{2\pi}{3}$ and ending at a point at infinity with $\frac{7\pi}{6} < \text{arg(t)} < \frac{3\pi}{2}$.

\textcolor{black}{For reason that will become clear shortly, it is convenient to rewrite expression \eqref{eq:amp} in the following form
\begin{equation}
\label{eq:amp1}
\begin{aligned}
    &F(\bm{Q},\bm{b},\bm{\xi},\eta, a, f_0) =\\ &f_0\int\limits_{0}^{\infty}(1+\eta s)\exp{-s\left(1+\frac{1}{4}Q^2a^2\right)}\Psi_{\perp}(b_x,b_y,s)ds.
\end{aligned}
\end{equation}
Notice that the integrand in \eqref{eq:amp1} taken at $s = 0$ is simply the packet's transverse wave function in real space with the impact parameter being the argument $\Psi_{\perp}(b_x,b_y) = \Psi_{\perp}(b_x,b_y,s = 0)$ and $|\Psi_{\perp}(x,y)|^2 = |\Psi_{\perp}(x,y,s=0)|^2$ is the packet's spatial distribution presented in Fig.\ref{Fig:1}}

\textcolor{black}{Had we considered a wave packet in momentum representation \eqref{eq:airy} with the transverse size
\begin{equation}
    \sigma_{\perp}'^2 = \sigma_{\perp}^2+\frac{a^2s'}{4} = \xi_{x,y}^2\rho^2_{x,y}(s'),
\end{equation}
and a nonzero average projections of the momentum
\begin{equation}
    \expv{k_{x,y}} = \frac{Q_{x,y}a^2s'}{4\sigma^{2'}_{\perp}} = \frac{1}{2\rho^2_{x,y}(s')}Re(\zeta_{x,y}(s'))
\end{equation}
the transverse wave function would have been exactly $\Psi_{\perp}(b_x,b_y, s = s')$. In the light of it, the amplitude for scattering of a wave-packet on a hydrogen-like atom (or Yukawa potential) is represented by superposition of the the wave-packets of the same shape, but with different parameters. In this interpretation, $\rho^2_{x,y}(s)$ is a dimensionless parameter determining the size of the packet and $\zeta_{x,y}(s)$ is a dimensionless parameter the real part of which determines the direction and the absolute value of the average transverse momentum of the wave-packet and the imaginary part determines its average position in space.}

\begin{figure*}[t]
\begin{minipage}[h]{0.32\linewidth}
\center{\includegraphics[height=4.5cm]{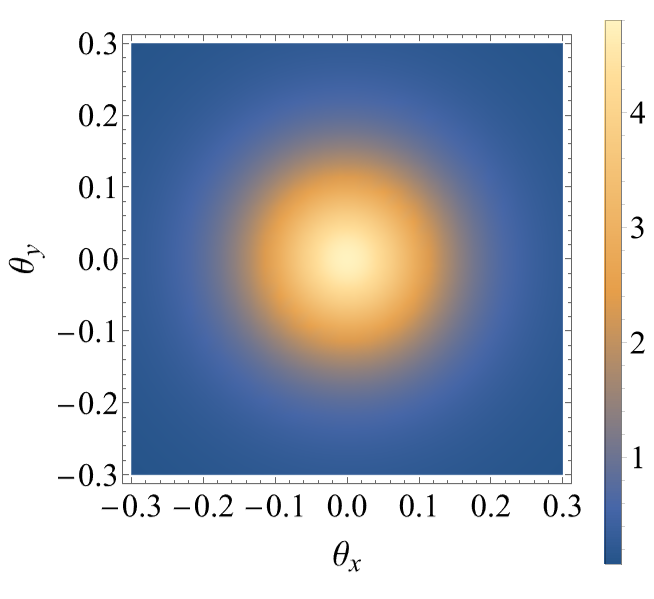} \\ a)
$\sigma_{\perp}=a, b_x=b_y = 0$.}
\end{minipage}
\begin{minipage}[h]{0.32\linewidth}
\center{\includegraphics[height=4.5cm]{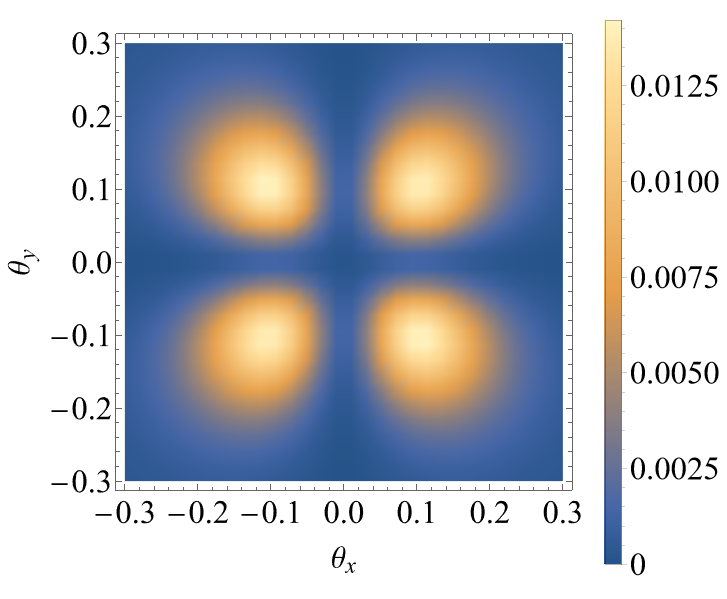} \\ b)
$\sigma_{\perp}=a, b_x=b_y\approx 4.8 \sigma_{\perp}$.}
\end{minipage}
\begin{minipage}[h]{0.32\linewidth}
\center{\includegraphics[height=4.5cm]{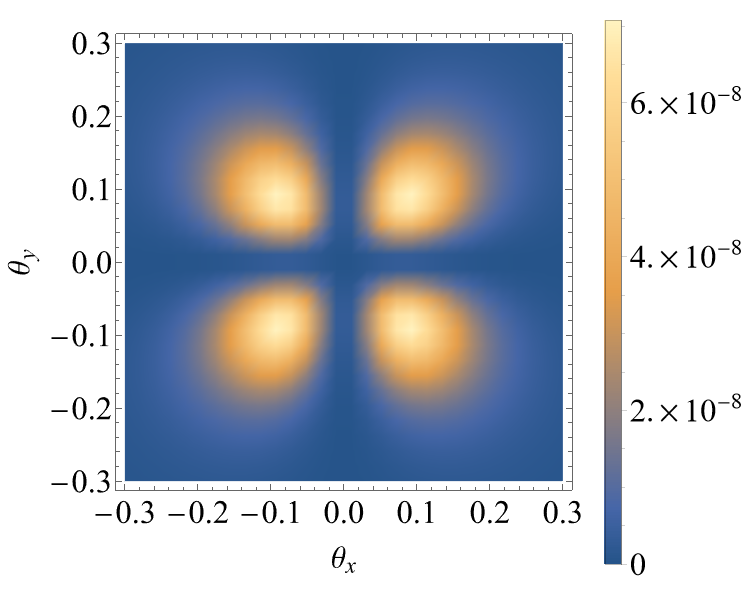} \\ c)
$\sigma_{\perp}=5a, b_x=b_y\approx 4.8 \sigma_{\perp}$.}
\end{minipage}
\begin{minipage}[h]{0.32\linewidth}
\center{\includegraphics[height=4.5cm]{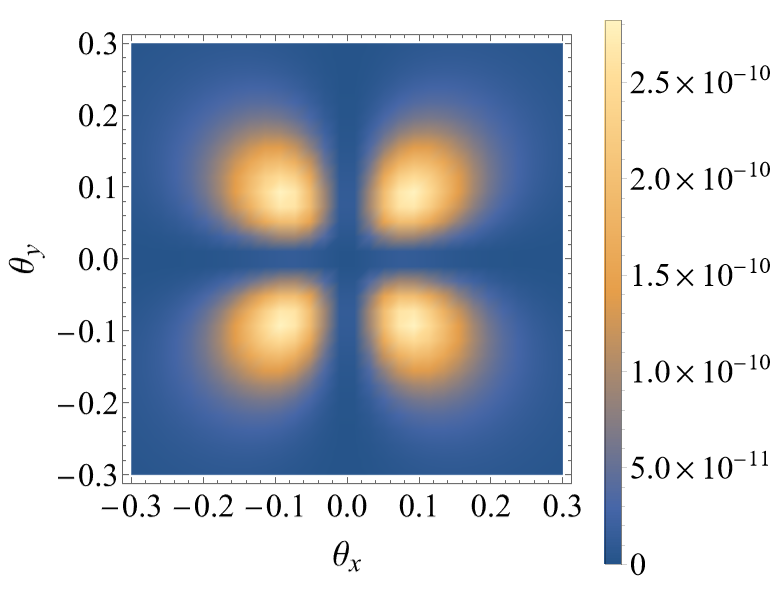} \\ d)
$\sigma_{\perp}=10a, b_x=b_y\approx 4.8 \sigma_{\perp}$.}
\end{minipage}
\begin{minipage}[h]{0.32\linewidth}
\center{\includegraphics[height=4.5cm]{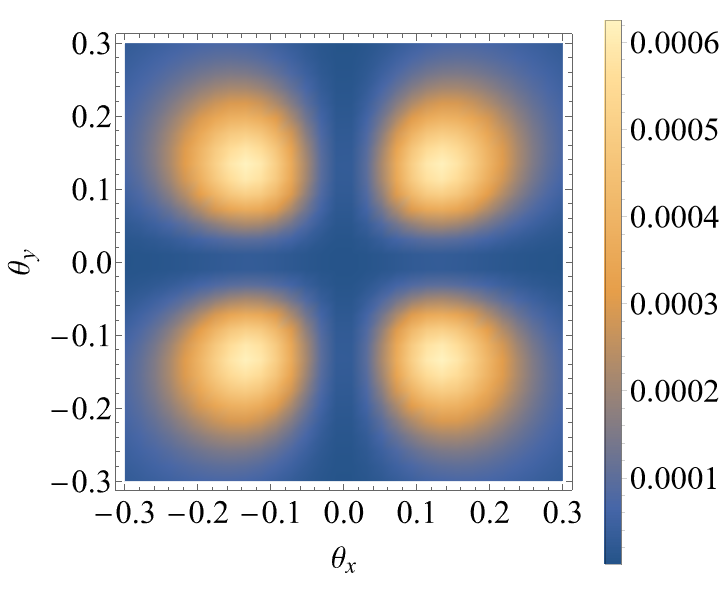} \\ e)
$\sigma_{\perp}=a, b_x=b_y\approx11.17\sigma_{\perp}$.
}
\end{minipage}
\caption{Scattering on a single atom: the dependence of the scattering probability density $\frac{d\nu}{d\Omega}(\theta_x,\theta_y)$ on the flat angles for the atom being located at different special points of the first type defined by \eqref{eq:fun} - first minima (a,c,d), third minima (b) for different transverse sizes of the packet - $\sigma_{\perp} = a, 5a, 10a$.}
\label{Fig:2}
\end{figure*}
\begin{figure}[h!]
\center{\includegraphics[width=1\linewidth]{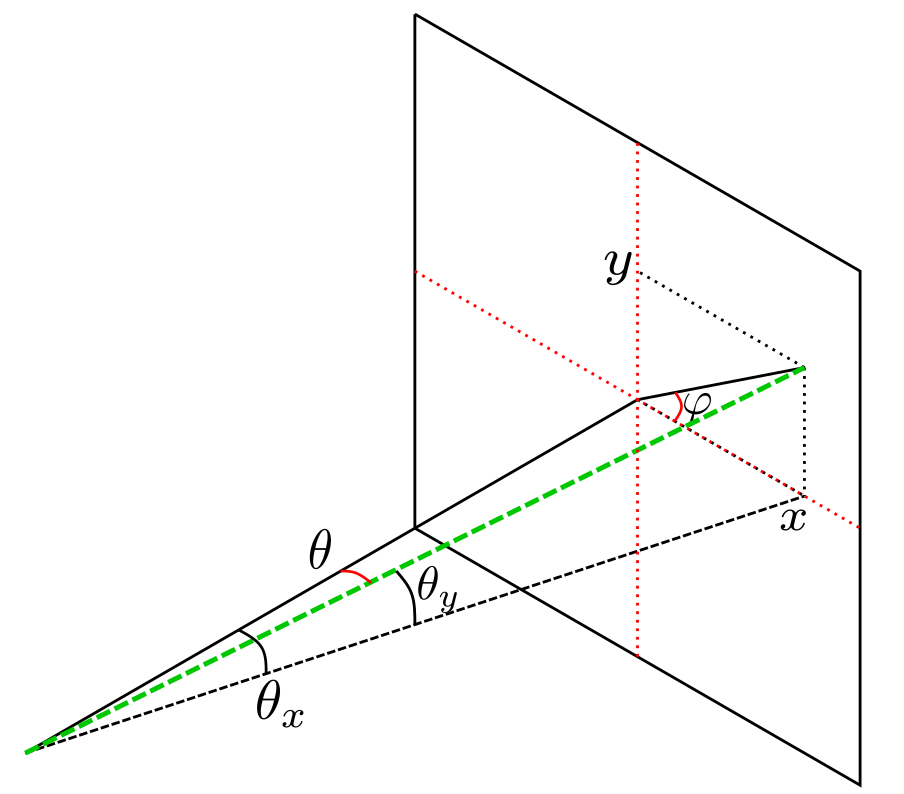}}
\caption{Angular distribution on flat angles $\theta_x$ and $\theta_y$ is almost equivalent to the distribution with respect to cartesian coordinates in the plane of detector.
$\cos \theta = \cos \theta_x \cos \theta_y; \:
\sin \varphi = \frac{\sin \theta_y}{\sqrt{1-\cos^2 \theta_x \cos^2 \theta_y}}$.}
\label{Fig:fa}
\end{figure}
\subsection{Scattering on a macroscopic target}
After the discussion of scattering by a single potential, let us
briefly describe scattering on a macroscopic (infinitely wide) target, which consists of randomly distributed potential centers. In this case we would need to integrate over all potential centers' positions and introduce the averaged cross-section as the integration of the number of events over all the impact parameters $\bm{b}$ and dividing the expression by the number of particles in the incident packet. The result is
\begin{equation}
\label{AvCS}
    \frac{d \Bar{\sigma}}{d\Omega} = \frac{1}{N_e}\int\frac{d \nu}{d\Omega} d^2 \bm{b}
\end{equation}
\textcolor{black}{The wave function of the packet approaching a target at impact parameter $\bm{b}$ can be written as 
\begin{equation}
    \Phi_{\perp}({\bm{k_{\perp}}}) = a(\bm{k_{\perp}})e^{-i\bm{k_{\perp}b}}
\end{equation}
Therefore, expression \eqref{AvCS} is proportional to the integral
\begin{equation}
    I = \int F(\bm{Q}) F^*(\bm{Q}) d^2 b,
\end{equation}
where 
\begin{equation}
   F(\bm{Q}) = \int f(\bm{Q} - \bm{k}_{\perp}) a(\bm{k_{\perp}})e^{-i\bm{k_{\perp}b}}\frac{d^2k_{\perp}}{(2\pi)^2}
\end{equation}
After the integration over $\bm{b}$ and $\bm{k_{\perp}}$ we obtain
\begin{equation}
\label{eq:av}
    \frac{d \Bar{\sigma}}{d\Omega} = \frac{1}{\cos \theta_k}\int |f(\bm{Q} - \bm{k}_{\perp})|^2 |\Phi_{\perp}(\bm{k}_{\perp})|^2d^2k_{\perp}
\end{equation}
(see details in \cite{Karlovets2015}).} Importantly, there is no dependence on the phase in expression \eqref{eq:av} meaning that scattering of the wave-packet  on a macroscopic target is merely defined by its transverse probability density and for the Airy packet \textit{it is the same as for the Gaussian one}.

\subsection{Scattering on a mesoscopic target}
In a more realistic experimental scenario a focused electron beam collides with a localized atomic target. In order to account for the geometrical effects in such a scenario we describe the target as an incoherent superposition of potential centers. The density of the scatterers in the transverse plane is characterized by a distribution function $n(\bm{b})$, which is normalized as follows:
\begin{equation}
\label{eq:dis}
    \int n(\bm{b}) d^2 \bm{b} = 1.
\end{equation}
For the numerical analysis below we take $n(\bm{b})$ to be a Gaussian function:
\begin{equation}
    n(\bm{b}) = \frac{1}{2\pi \sigma_b^2}\exp (-\frac{(\bm{b}-\bm{b_0})^2}{2\sigma_b^2}).
\end{equation}
For such a scenario the number of events compared to \eqref{eq:NoE} modifies in the following way:
\begin{equation}
    \frac{d\nu}{d\Omega} = \frac{N_e}{\cos\theta_k}\int|F(\bm{Q}, \bm{b})|^2 n(\bm{b}) d^2 \bm{b}.
\end{equation}

\begin{figure*}[t]
\begin{minipage}[h]{0.32\linewidth}
\center{\includegraphics[height=4.5cm]{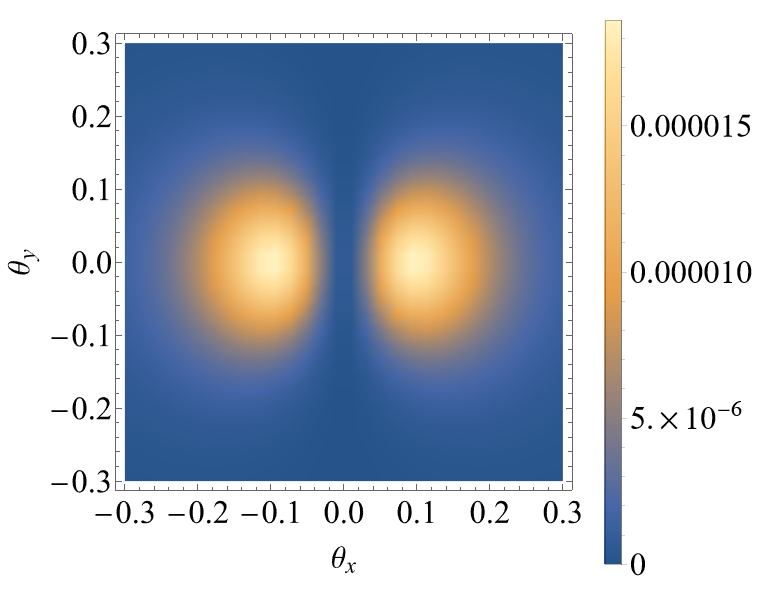} \\ a)$\sigma_{\perp} = 5a, b_x\approx 4.8 \sigma_{\perp}, b_y=0$.}
\end{minipage}
\begin{minipage}[h]{0.32\linewidth}
\center{\includegraphics[height=4.5cm]{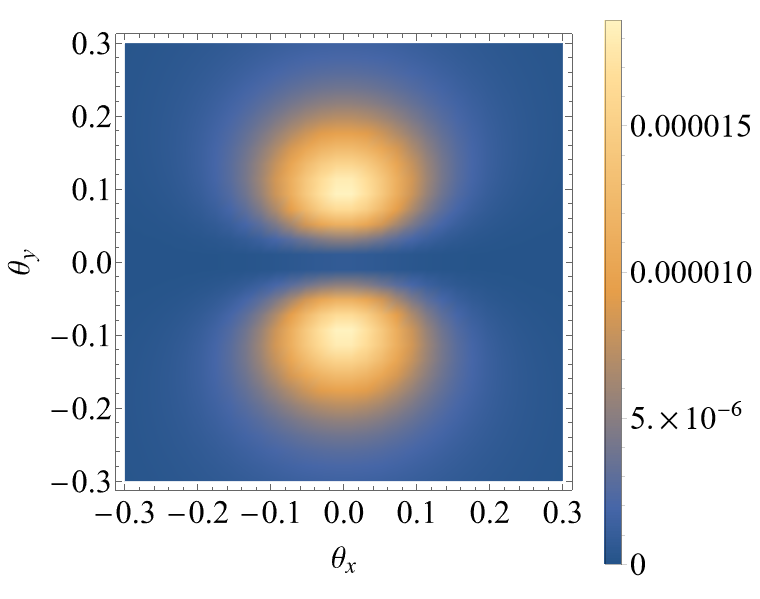} \\ b)$\sigma_{\perp}=5a, b_y\approx 4.8 \sigma_{\perp}, b_x=0$.}
\end{minipage}
\begin{minipage}[h]{0.32\linewidth}
\center{\includegraphics[height=4.5cm]{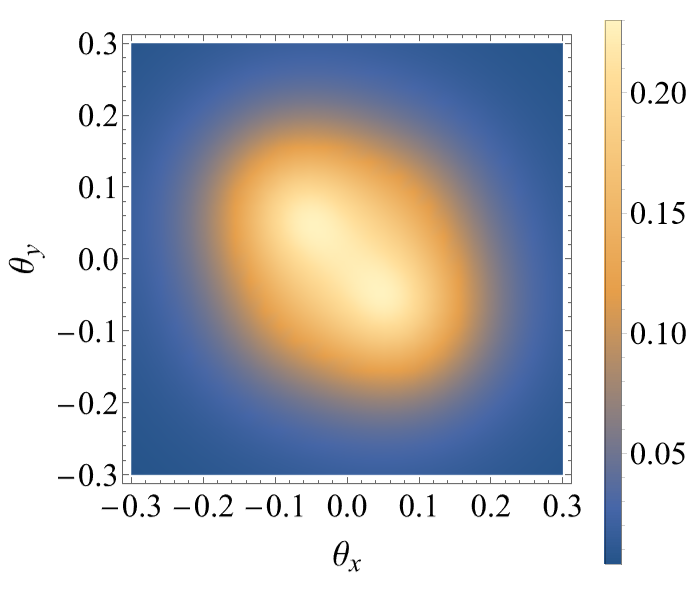} \\c)
$\sigma_{\perp}=a, b_x=b_y = 4 \sigma_{\perp}$.}
\end{minipage}
\caption{Scattering on a single atom: the dependence of the scattering probability density $\frac{d\nu}{d\Omega}(\theta_x,\theta_y)$ on the flat angles for the atom being located at different special points of the second type for the packet's transverse size $\sigma_{\perp} = 5a$ (a,b) and a transitional point and the packet's transverse size $\sigma_{\perp} = a$ (c).}
\label{Fig:3}
\end{figure*}
\section{Results}

All the figures in the following subsections are presented for $p_ia = p_fa = 10$. \cite{Karlovets2015}. \textcolor{black}{Such a momentum value corresponds to non-relativistic, but still ``fast'' electrons. It can be seen from the expression \eqref{eq:amp} that increasing the projectile momentum simply leads to faster exponential decay of the scattering amplitude as a function of the polar angle $\theta$, i.e. ``narrowing'' of the scattering pattern.} For the scattering on a hydrogen atom this corresponds to the kinetic energy $\varepsilon_i = 1.36$ keV. We also assume that $\xi_{x, y} =2\sigma_{\perp}$  unless stated otherwise \cite{Karlovets2017a, VolochBloch2013}. We normalize the packet's wave function to 1, meaning that only one particle collides the target. Thus, in the following sections it is more correct to speak of \textit{scattering probability density} rather than of the number of events.

\subsection{Scattering pattern for central collision}

The final results are more illustrative when expressed via the so-called flat angles \eqref{eq:flat} \cite{Ivanov2013} presented in Fig.\ref{Fig:fa}
\begin{equation}
\label{eq:flat}
    \cos \theta = \cos \theta_x \cos \theta_y,
\sin \phi = \frac{\sin \theta_y}{\sqrt{1-\cos^2 \theta_x \cos^2 \theta_y}}.
\end{equation}

These angles are connected with the Cartesian coordinates on a surface of a flat detector.

The number of events for scattering on a single atom in reality turns out to be very sensitive to the relative position of the atom and the wave-packet's probability density minima.
As can be seen in Fig.(\ref{Fig:2},a), the scattering pattern looks rather symmetric for the head-on collision, when the atom is placed on the axis of the packet's propagation with the impact parameters being $b_x = b_y = 0$. This can be interpreted in the following way: the first maxima of the Airy packet is rather wide, about $5\sigma_{\perp} \approx 5a$, which is why with the potential's radius of action being equal to $a$ the atom only feels the vicinity of the first maximum and does not feel the Airy nature of the wave packet which manifests itself in oscillatory behaviour in one quadrant of the $(x,y)$ - plane. In a simple approximation we could say that the atom feels the packet as a Gaussian one and the pattern then turns out to be a symmetric circle \textcolor{black}{as expected for a such case \cite{Karlovets2015}}.

\begin{figure*}[t]
\begin{minipage}[h]{0.32\linewidth}
\center{\includegraphics[height=4.5cm]{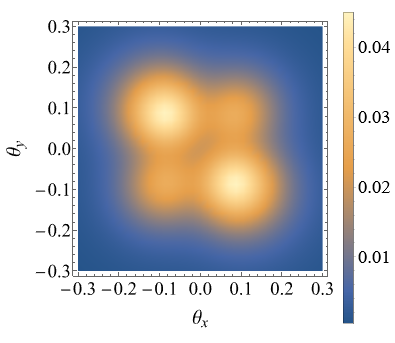} \\ a)$b_{x, y} = 4.4\sigma_{\perp}$.}
\end{minipage}
\begin{minipage}[h]{0.32\linewidth}
\center{\includegraphics[height=4.5cm]{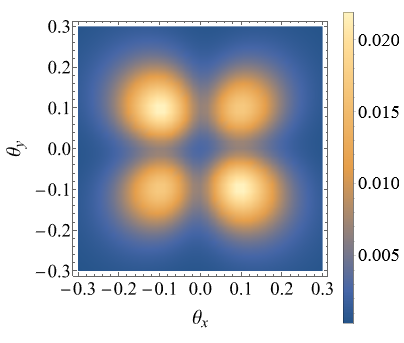} \\ b)$b_{x, y} = 4.6\sigma_{\perp}$.}
\end{minipage}
\begin{minipage}[h]{0.32\linewidth}
\center{\includegraphics[height=4.5cm]{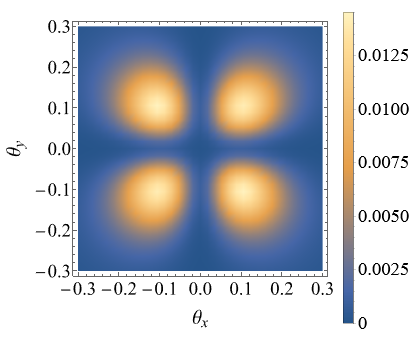} \\c)$b_{x, y} = 4.8\sigma_{\perp}$.}
\end{minipage}
\begin{minipage}[h]{0.32\linewidth}
\center{\includegraphics[height=4.5cm]{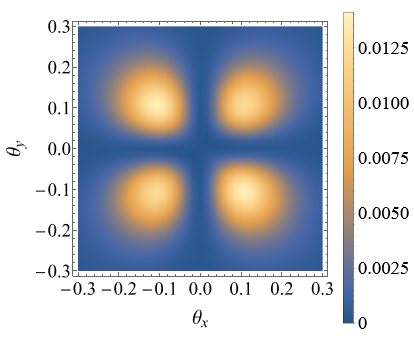} \\ d)$b_{x, y} = 5\sigma_{\perp}$.}
\end{minipage}
\begin{minipage}[h]{0.32\linewidth}
\center{\includegraphics[height=4.5cm]{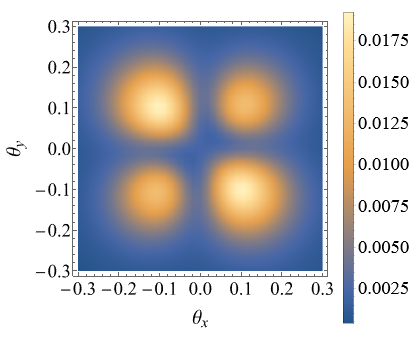} \\ e)$b_{x, y} = 5.2\sigma_{\perp}$.}
\end{minipage}
\begin{minipage}[h]{0.32\linewidth}
\center{\includegraphics[height=4.5cm]{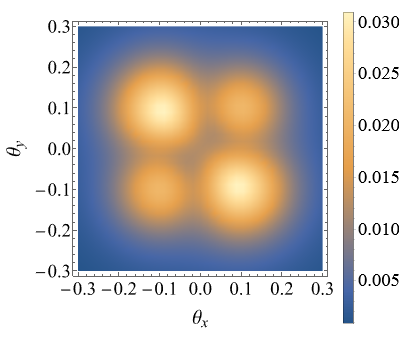} \\f) $b_{x, y} = 5.4\sigma_{\perp}$.}
\end{minipage}
\caption{Scattering on a single atom: comparison of scattering patterns within the vicinity of a special point of the first type for evaluating the sensitivity with respect to inaccuracy of impact parameter value for packet size $\sigma_{\perp} = 2a$}
\label{Fig:7}
\end{figure*}

\subsection{Scattering pattern for special points of type 1}
However, there are totally different scattering patterns that occur when the atom is placed in the minima of the Airy packet's probability density. These points are defined conditions \eqref{eq:fun} that are both satisfied \textit{simultaneously}:
\begin{equation}
\label{eq:fun}
\begin{aligned}
    &\operatorname{Ai}\left(-\frac{b_x}{\xi_{x}}+\frac{\sigma_{\perp}^4}{\xi_{x}^4}\right) = 0,\\ &\operatorname{Ai}\left(-\frac{b_y}{\xi_{y}}+\frac{\sigma_{\perp}^4}{\xi_{y}^4}\right) = 0.
\end{aligned}
\end{equation}
In the following discussion we will refer to such points as special points \textit{of the first type}. For special points circular scattering pattern is replaced by a 4-petal pattern.

As can be seen in Fig.\ref{Fig:2} the probability of the process vanishes for the forward scattering, which corresponds to the flat angles being in the vicinity of zero. In Fig.(\ref{Fig:2},b) we illustrate the scattering probability density for scattering of a narrow packet when its transverse size is equal to the potential's radius of action ($\sigma_{\perp} = a$, e.g, Bohr radius in case of hydrogen atom) on the atom that is placed in the first minima of the transverse probability density ($b_x \approx 4.8\sigma_{\perp}$) and see that the probability density is of order $10^{-3}-10^{-2}$. Increasing the wave-packet's transverse size leads to a relatively sharp decrease in the magnitude of the probability density, e.g. increasing the packet's size by a factor of 5 leads to a decrease in the probability density of the order of $10^{-7}$, nevertheless, the scattering pattern qualitatively remains the same.

Switching the position of the atom from the first minima to a different one keeps the scattering pattern visually the same, however, its the magnitude decreases. As can be seen in Fig.(\ref{Fig:2},e) when the atom is placed in the \textit{third} minima with $b_x = b_y \approx 11.17\sigma_{\perp}$ the probability density decreases by the order of $10^{-2}$. What else can be noticed is that figures (\ref{Fig:2},b),(\ref{Fig:2},c),(\ref{Fig:2},d) are visually the same and the difference is only in the values of the probability density, however, for a different minima Fig.(\ref{Fig:2},e) the petals are spaced further apart from each other.

\subsection{Scattering pattern for special point of type 2}
\begin{figure*}[t]
\begin{minipage}[h]{0.42\linewidth}
\center{\includegraphics[width=1\linewidth]{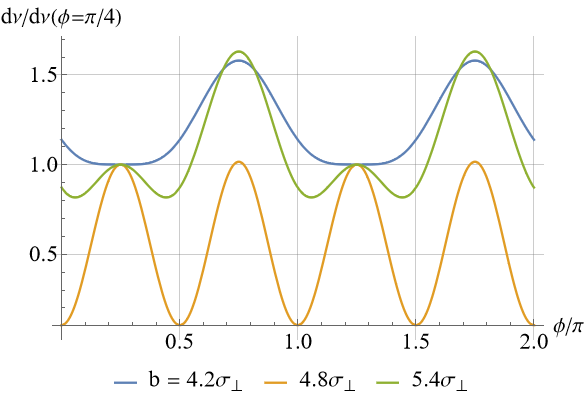} \\ a) $\sigma_{\perp}=a$}
\end{minipage}
\begin{minipage}[h]{0.42\linewidth}
\center{\includegraphics[width=1\linewidth]{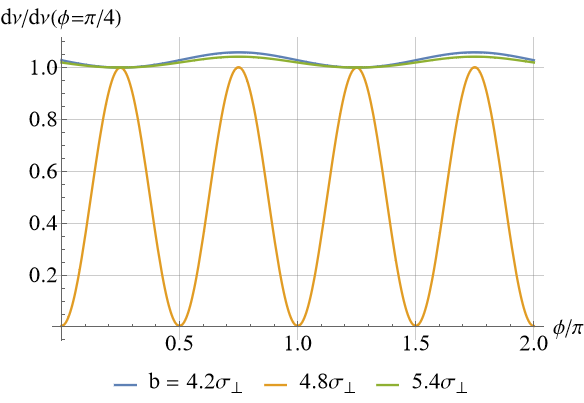} \\ b)
$\sigma_{\perp}=5a$}
\end{minipage}
\begin{minipage}[h]{0.3\linewidth}
\end{minipage}
\begin{minipage}[h]{0.42\linewidth}
\center{\includegraphics[width=1\linewidth]{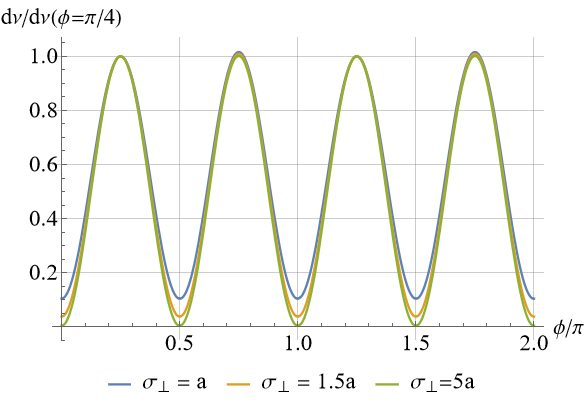} \\ c) $b_{x, y}\approx 4.8\sigma_{\perp}$}
\end{minipage}
\begin{minipage}[h]{0.42\linewidth}
\center{\includegraphics[width=1\linewidth]{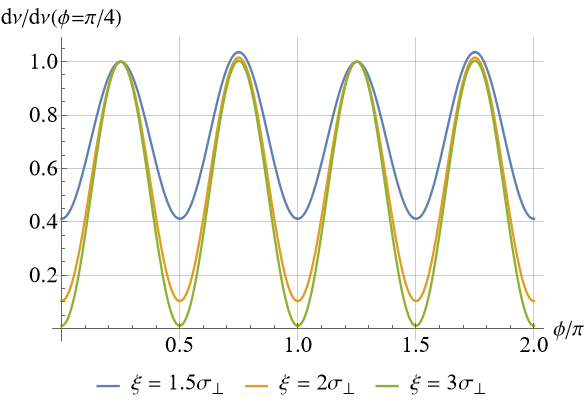} \\ d) $\sigma_{\perp} = a, \:\operatorname{Ai}\left(-b_{x, y}/ \xi_{x,y}+\sigma_{\perp}^4/\xi_{x_y}^4\right) = 0$}
\end{minipage}
\begin{minipage}[h]{0.3\linewidth}
\end{minipage}
\caption{\textcolor{black}{Azimuthal dependence of the the ratio of probabilities} $d \nu (\theta, \varphi) / d \nu (\theta, \varphi=\pi/4)$ for different values of impact parameter, packet's size and parameters $\xi_{x, y}$. We assume that $b_x=b_y$ and that outgoing electrons are detected at the polar angle $\theta=0.1$ rad.}
\label{Fig:5}
\end{figure*}

In reality the probability density is at its minima when either one of the conditions \eqref{eq:fun} is fulfilled, thus one could ponder what would happen to the scattering pattern if the atom is placed in a position that satisfies \textit{only one of them}. We will refer to such points as special points of \textit{the second type}. We illustrate such scattering patterns when the atom's position satisfies the condition of only the first or the second Airy function in \eqref{eq:fun} being equal to zero in Fig.(\ref{Fig:3},a),(\ref{Fig:3},b) respectively. As can be seen in this case the circular scattering pattern from Fig.(\ref{Fig:2},a) splits into a 2-petal pattern in two orthogonal directions depending on which Airy function vanishes at that point. Similar to the case of placing the atom in a special point of the first type the probability density $\frac{d\nu}{d\Omega}(\theta_x,\theta_y)$ vanishes for very small flat angles meaning vanishing probability of forward scattering.

Combining observations of the scattering from special point of the first and of the second type we can deduce that each of the two equations in \eqref{eq:fun} behaves as a splitting mechanism that transforms circular-like parts of the scattering patterns into two pieces.

Now, we could try to go beyond the cases considered thus far and study what would happen to the scattering pattern \textit{in the vicinity of the special point}. In Fig.(\ref{Fig:3},c) we illustrate the case when impact parameter $b_x = b_y = 4\sigma_{\perp}$ is close to the special point of the first type yet none of the conditions \eqref{eq:fun} is satisfied. For such a scenario we notice transitional behaviour of the scattering pattern and will refer to such points as \textit{transitional}.
\textcolor{black}{Note that such a pattern also resembles the scattering of a Gaussian wave-packet with nonzero impact parameter \cite{Karlovets2015}. We could interpret this result similarly to the central collision: the contribution of the main maximum in this case significantly exceeds the contribution of other maxima.}

\subsection{Feasibility of scattering patterns and their sensitivity to precision of atom's placement}
In Sec. III B. the density plots are given for impact parameters defined up to $0.1\sigma_{\perp}$. For narrow packets with $\sigma_{\perp} = a$ this implies the necessity to control the atoms position with the degree of precision of $0.01 - 0.1 $\AA, which is hardly feasible nowadays. For larger packets with $\sigma_{\perp} = 5a, 10a$ the required precision is insignificantly lower but from experimental point of view it is still impossible to achieve such accuracy and that is why we analyze the sensitivity of the scattering patterns with respect to inaccuracy of atom's position. In Fig.\ref{Fig:7} we present the series of scattering patterns for values of impact parameter being gradually increased by $0.2\sigma_{\perp}$. We see that for the values $|\bm{b}| = 4.6\sigma_{\perp},4.8\sigma_{\perp},5\sigma_{\perp},5.2\sigma_{\perp},6\sigma_{\perp}$ which are all in the vicinity of the special point of the first type the scattering pattern deviates from the one corresponding to atom's placement precisely the minima of the probability density to a degree depending on the value of impact parameter, and yet the 4-petal form is still more or less recognizable, thus we can roughly estimate the required accuracy as $\sim 0.6 \sigma_{\perp} = 1.2 a$. Moreover, for \textcolor{black}{decreasing} the required accuracy of atom's placement one could consider scattering on a Rydberg atom with the radius of action $\tilde{a} = a n^2$ which is much greater than $a$ for large $n$. This would lead to rescaling of the whole problem keeping all the expressions the same and the required accuracy could then be of the order $1.2\tilde{a} \sim a n^2 \sim 10$ nm for $n = 10$.

\subsection{Azimuthal dependence for single atom
scattering}
After discussing the scattering patterns, let us dig deeper into the scattering process for different scenarios. In reality the number of events is a sharp function in terms of the polar angle and it grows sharper with the increase of the momentum which we put equal $p_i a = p_f a =10$ as stated before. Hence we fix the polar angle to be $\theta = 0.1$ rad and study the azimuthal dependence of the number of events for scattering on a single atom.

In Fig(\ref{Fig:5},a) we display the normalized number of events for scattering on an atom placed at the special point of the first type ($b = 4.8\sigma_{\perp}$ - orange line) and at two transitional points ($b = 4.2\sigma_{\perp}, b = 5.4\sigma_{\perp}$). In the curve describing special point we see 4 visibly equivalent maxima that correspond to the 4-petal pattern. Shifting the impact parameter, two of the maxima flatten out and for $b = 4.2\sigma_{\perp}$ there are only 2 visible peaks like in a scattering pattern Fig.(\ref{Fig:3},c) where the impact parameter is taken to be $b = 4\sigma_{\perp}$.

Increasing the size of the packet allows one to discover that the scattering pattern for the special point is in general much more sensitive to the azimuthal angle as can be seen in Fig(\ref{Fig:5},b) as for the size of the packet $\sigma_{\perp} = 5a$ the two curves for the transitional points (blue line for $|b| = 4.2\sigma_{\perp}$, black line for $|b| = 5.4\sigma_{\perp}$) flatten out compared to the curve describing the special point of the first type.

If we now compare the azimuthal dependence for special points with different sizes of the packet Fig.(\ref{Fig:5},c), we see that the narrower the wave-packet is, the smaller is the \textcolor{black}{ratio} of number of events as a function of the azimuthal angle and the greater is its minimum value.

Finally, we consider scattering on an atom placed at a special point and study how the dependence changes when we change the phase of the packet meaning different $\xi_x,\xi_y$ Fig.(\ref{Fig:5},d). The pattern of the dependence for different $\xi_x,\xi_y$ remains the same yet the \textcolor{black}{ratio of probabilities} decreases. The idea behind it is that for smaller values of $\xi_x,\xi_y$ the wave-packet is less and less distinguished as an Airy packet and starts to resemble the Gaussian one more and more and this leads to the flattening of the azimuthal dependence of the number of events.

\subsection{Scattering pattern for a mesoscopic target}
\begin{figure}[h]
\center{\includegraphics[width=1\linewidth]{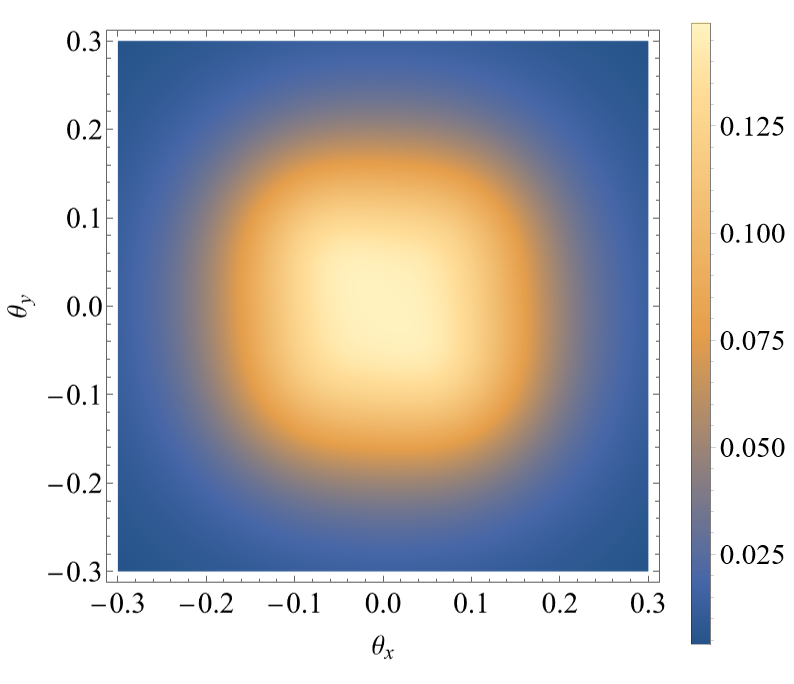}}
\caption{Scattering on a mesoscopic target: the dependence of the scattering probability density $\frac{d\nu}{d\Omega}(\theta_x,\theta_y)$ on the flat angles for scattering of the Airy packet with $\sigma_{\perp} = a$ on a mesoscopic target centered at the special point of the first type \textcolor{black}{($b_{x, y} \approx 4.8 \sigma_{\perp}$)} with the width $\sigma_{b} = a$. This corresponds to the case of a target consisting of \textit{very few atoms}.}
\label{Fig:8}
\end{figure}
In this section we turn to a more realistic experimental scenario where we investigate scattering on a mesocopic target of finite size. To study such a process we describe the target with the distribution function \eqref{eq:dis}. In the case of incoherent scattering on a mesoscopic target, meaning that we average the scattering probability density rather than the scattering amplitude, we can think of the scattering pattern as an overlay of different scattering patterns for different placings of a single atom. Thus, we will take into account a continuum of points for atom's placings that are neither special points of the first nor of the second type the scattering pattern for which is more or less circular and scattering pattern from special points with 4 or 2-petals and also the transitional cases. As a result, we expect the 4-petal and 2-petal forms to smear out and end up with a pattern which is non-vanishing for forward scattering. 

In Fig.(\ref{Fig:8}) we present the scattering pattern for the transverse size of the beam and the width of the target being equal $\sigma_{\perp} = \sigma_{b} = a$ and the center of the target located at the special point of the first type. As can be seen in the figure, there are no separated peaks as expected, yet after averaging over impact parameters the pattern looks "quadratic" rather than azimuthally symmetric even with a naked eye. The reason for it is the narrow width of the target. As discussed above the resulting scattering pattern can be thought of as an overlay of scattering patterns on the atom the position of which is defined by the impact parameter $b_x = b_y = b$ with $b$ passing all the values in the interval $(4.8\sigma_{\perp}-\sigma_{b}, 4.8\sigma_{\perp} + \sigma_{b}) = (3.8a,5,8a)$ which does not contain the point corresponding to the position of the main maximum of the probability density ($|\bm{b}| \approx 2.16a$) of the Airy packet the amplitude of the scattering pattern for which Fig(\ref{Fig:2},a) is of several orders higher. Thus the effects of the internal structure of the packet are still visible. However from the experimental point of view, the size of target $\sigma_{b} = a$ is hardly achievable and the more realistic scenario would imply $\sigma_{b} \sim 10-20 a$. For such a target the interval of the values of the impact parameters of overlaying patterns would inevitably contain the main maxima and azimuthal asymmetry would be practically invisible. 

For a wave-packet with $\xi_x = \xi_y = \xi$ and the position of the center of the target $b_0$ being one of the special points of the first type the following \textcolor{black}{semi-empirical} inequality
\begin{equation}
\label{eq:Size}
    -\frac{b}{\xi} + \frac{\sigma_{\perp}^4}{\xi^4} < -1.018
\end{equation}
has to be satisfied for all $b \in (b_0 - \sigma_b,b_0+\sigma_b)$ for the observation of the contribution of the internal structure of the wave-packet to be possible. The inequality simply states that for all absolute values of the impact parameter that contribute to the resulting scattering pattern the argument of the Airy function in \eqref{eq:fun} should be less than the position of the main maxima ($b \approx -1.018$), i.e., the main maxima does not contribute to the pattern. With the use of \eqref{eq:Size} we could estimate the critical width of the target $\sigma_c$ that could allow observation of the internal degrees of freedom of the packet in the scattering pattern as the value that satisfies
\begin{equation}
\begin{aligned}
    -\frac{b_0 - \sigma_c}{\xi} + \frac{\sigma_{\perp}^4}{\xi^4} = -1.018,\\
    \sigma_c = b_0 - \frac{\sigma_{\perp}^4}{\xi^3} - 1.018\xi.
\end{aligned}
\end{equation}
When $b_0 = 4.8\sigma_{\perp}$, which corresponds to the first special point, and $\xi = 2\sigma_{\perp}$ the critical size estimate is $\sigma_c = 2.64\sigma_{\perp}$
From expression \eqref{eq:Size} it is clear that to achieve visible asymmetry in scattering patterns on reasonable-sized mesoscopic targets one should place them at a father special points of the first type. We remind that there are infinitely many special points of the first type as a result of oscillatory behaviour of the Airy function with the first three described by $b \approx 4.8\sigma_{\perp}, b \approx 8.25\sigma_{\perp}, b \approx 11.16\sigma_{\perp}$.

Nonetheless, for the fixed polar angle $\theta = 0.1$ rad the variation of the\textcolor{black}{ratio of probability densities $d \nu (\theta, \varphi) / d \nu (\theta, \varphi=\pi/4)$} as of the function of the azimuthal angle for $\sigma_{\perp} = \sigma_{b} = a$ is already of the order of $0.05$. Increasing the size of the target inevitably leads to more and more azimuthally symmetric (circular) scattering patterns as in the limit of a macroscopic target the cross-section of the process no longer depends on the phase of the wave-packet, as discussed in Sec.II D, and for $\sigma_b \gtrsim \sigma_{c} = 2.64\sigma_{\perp} = 2.64a$ the asymmetry vanishes. For the target width $\sigma_b = 10a$ the asymuthal variation of the \textcolor{black}{ratio of probabilities} is already of the order of $10^{-3}$ and for $\sigma_b = 30$ it decreases to $10^{-4}$.

\section{Discussion and conclusion}
\textcolor{black}{
Recalling the definition for the plane-wave scattering amplitude entering expression \eqref{eq:NoE} one can arrive at the following expression for the non-plane wave amplitude 
\begin{equation}
    F(\bm{Q}) = -\frac{m_e}{2\pi}\int U(r)\Phi(r_{\perp})e^{-i\bm{Q}\cdot\bm{r}}d^3r,
\end{equation}
which allows one to retrieve information about the transverse wave function of the initial state by applying Fourier transform to derive
\begin{equation}
\label{eq:in}
    \Phi(r_{\perp}) = -\frac{U^{-1}(r)}{4m_e\pi^2}\int F(\bm{q})e^{i\bm{q}\cdot\bm{r}}d^3q,
\end{equation}
which can be rewritten in the following form
\begin{equation}
\label{eq:inverse}
    \Phi(r_{\perp}) = -\frac{m_e}{2\pi}\frac{\int F(\bm{q})e^{i\bm{q}\cdot\bm{r}}d^3q}{\int f(\bm{q})e^{i\bm{q}\cdot\bm{r}}d^3q}.
\end{equation}
Technically, expression \eqref{eq:inverse} solves the inverse scattering problem as it allows one to derive the wave function of scattered particle from the scattering amplitude. However, we stress that a fully satisfying result would be the ability to retrieve at least a portion of information about the wave-packet's distribution solely from the absolute value of the scattering amplitude as experimental measurements can in reality be insensitive to the phase of the scattering amplitude in a majority of scenarios.}

\textcolor{black}{In summary, we have applied the generalized Born approximation to the problem of elastic scattering of a non-relativistic charged Airy wave-packet on a potential field.} For single hydrogen atom the probability density sharply decreases with the increase of the packet's width. In particular, as it changes from $\sigma_{\perp} = a = 0.5\text{\AA}$ to $\sigma = 5a = 2.5\text{\AA}$ the values drop from $0.01$ to $10^{-8}$. For the angular dependence we have found that the probability density $d\nu/d\Omega(\theta_x,\theta_y)$ as a function of flat angles \eqref{eq:flat} acquires the 4-petal shape
when the atom is placed at the special points of \textit{the first type} and the 2-petal shape when placed at the  special points of \textit{the second type}. These points correspond to the packet's density minimum, where 4-petal shapes correspond to satisfying both conditions in \eqref{eq:fun} simultaneously and 2-petal shapes occur when only one of them is satisfied.
Characteristic scattering patterns allows one to get insight into the internal structure of the wave-packet by placing the atom at different points, i.e., performing \textit{tomography of the wave-packet} using the atom as a probing tool. 

In reality, such specific placings of the atom require the precision of the atom's position of the order of $0.1\sigma_{\perp} \sim 0.1a-1a \sim 0.1-1$\AA. From experimental point of view it seems to be vary hard to control the position of the atom with such accuracy. 
That is why we have introduced the idea of using Rydberg atoms, which have much larger radius of action $\tilde{a} = an^2 \sim 10^{-8} - 10^{-4}$ m, where $n$ is a principal quantum number. This simply leads to rescaling of the whole problem keeping all the expression and results intact. Thus, one can study scattering of wave-packets with significantly larger sizes up to $\sigma_{\perp} \sim 10-100 \;\mu m$ and greatly decrease the required precision of atom's placement to micrometer scale since for Rydberg atoms $n$ can be as high as 1000 \cite{Aasen2022}.

Analyzing the \textcolor{black}{azimuthal dependence of the ratio of probabilities $d \nu (\theta, \varphi) / d \nu (\theta, \varphi=\pi/4)$} we have shown that it is in general much more sensitive to the azimuthal angle when the atom is placed at the special point of the first type. For $b=4.8\sigma_{\perp}$, which corresponds to the first special point, the \textcolor{black}{ratio variation} is of order of 1, whereas for $b$ shifted by $0.6\sigma_{\perp}$ it is roughly ten times smaller. This may turn out to be useful in different microscopy problems when the information regarding some object is extracted from azimuthal dependence of scattering patterns as setting the investigated object at the special point  would result in a more explicit behaviour.

Finally, we have discussed the most experimentally achievable  scenario, i.e., scattering on a mesoscopic target. For a toy-model case of a narrow packet and an equally narrow target the scattering pattern still contains information about the internal structure of the wave-packet, which is manifested in "quadratic" shape of the probability density. We have introduced the critical size of target as the maximum size, which could enable observation of the asymmetry in the scattering on a mesoscopic target, which for $\xi_x = \xi_y = 2\sigma_{\perp}, b_x = b_y = 4.8\sigma_{\perp}, \sigma_{\perp}  = a$ is \begin{equation}
    \sigma_c = 2.64a \;(\text{FWHM} = 6.22a).
\end{equation}
From analysis of the general expression for $\sigma_c$ it follows that to make the observation of azimuthal asymmetry more feasible apart from considering Rydberg atoms rather than the hydrogen atom one could place the target's center at a special point remote from the packet's main peak.

The nature of the studied effects and the problems that arise along the way are not unique for Airy beams. They are the result of non-Gaussian profiles and appear due to the presence of a phase $\varphi(\bm{p})$, where $\bm{p}$ is the particle's momentum. Similar scenarios could be studied for different manifestations of phases such as vortex states, Pearcy beams and their various generalizations \cite{He2022, Huang2022}.

\section{Acknowledgements}
We are grateful to S. Baturin, A. Volotka, V. Ivanov, A. Katanaeva, G. Sizykh, and A. Surzhykov for fruitful discussions
and criticism. The studies in Sec.II are supported by the Russian Science Foundation (Project No. 21-42-04412). The studies in Sec.III.A,B are supported by the Government of the Russian Federation through the ITMO Fellowship and Professorship Program. The studies in Sec.III.C,D are supported by the Ministry of Science and Higher Education of the Russian Federation (agreement no. 075-15-2021-1349). The work in Sec.III.E,F by D. Karlovets and D. Grosman was supported by the Foundation for the Advancement of Theoretical Physics and Mathematics “BASIS”. 

\newpage
\bibliographystyle{apsrev}
\bibliography{AiryLiterature}
\end{document}